\documentclass{ws-procs975x65}

\pdfoutput=1

\begin{document}

\title{GAME: GRB AND ALL-SKY MONITOR EXPERIMENT}

\author{LORENZO AMATI$^*$, RICCARDO CAMPANA, YURI EVANGELISTA, MARCO FEROCI, \\
FABIO FUSCHINO,  
CLAUDIO LABANTI, RUBEN SALVATERRA, GIULIA STRATTA and GIANPIERO TAGLIAFERRI}

\address{Italian National Institute for Astrophysics (INAF), \\
Italy\\
$^*$E-mail: amati@iasfbo.inaf.it
}

\author{FILIPPO FRONTERA, CRISTIANO GUIDORZI, PIERO ROSATI and LEV TITARCHUK}

\address{University of Ferrara, Department of Physics and Earth sciences,\\
Ferrara, Italy
}

\author{JO\~AO BRAGA and ANA PENACCHIONI}

\address{Instituto Nacional de Pesquisas Espaciais (INPE),\\
Sao Jose dos Campos, Brasil
}

\author{REMO RUFFINI and LUCA IZZO}

\address{International Center for Relativistic Astrophysics Network (ICRANet),\\
Pescara, Italy
}

\author{NICOLA ZAMPA and ANDREA VACCHI}

\address{INFN - Sezione di Trieste, \\
Trieste, Italy
}

\author{ANDREA SANTANGELO}

\address{University of Tubingen, \\
Tubingen, Germany 
}

\author{RENE HUDEC}

\address{
Czech Technical University in Prague, Faculty of Electrical Engineering \\ \&
Astronomical Institute of the Academy of Sciences of the Czech Republic, \\
Ondrejov
Czech Repubblic 
}

\author{ANDREJA GOMBOC and TOMAZ RODIC}

\address{Ljubljana University and SPACE-SL, \\
Ljubljana, Slovenia 
}

\author{on behalf of the GAME collaboration}

\address{http://www.iasfbo.inaf.it/$\sim$amati/game.pdf \\
}

\begin{abstract} 

We describe the GRB and All-sky Monitor Experiment (GAME) mission submitted by a 
large international collaboration (Italy, Germany, Czech Repubblic, Slovenia, 
Brazil) in response to the 2012 ESA call for a small mission opportunity for a 
launch in 2017 and presently under further investigation 
%in view of next 
for subsequent
opportunities. 
The general scientific objective is to perform measurements of key 
importance for GRB science and to provide the wide astrophysical community of an 
advanced X-ray all-sky monitoring system. The proposed payload was based on 
silicon drift detectors ($\sim$1--50 keV), CdZnTe (CZT) detectors ($\sim$15--200 
keV) and crystal scintillators in phoswich (NaI/CsI) configuration 
($\sim$20 keV--20 MeV), three well established technologies, for a total weight of $\sim$250 kg 
and a required power of $\sim$240 W. Such instrumentation allows a unique, 
unprecedented and very powerful combination of large field of view % FOV
(3--4 sr), a broad energy 
energy band extending from $\sim$1 keV up to $\sim$20 MeV, an energy resolution as 
good as $\sim$250 eV in the 1--30 keV energy range, a source location accuracy of 
$\sim$1 arcmin. The mission profile included a launch (e.g., by Vega) into a 
low Earth %LEO 
orbit, a baseline sky scanning mode plus pointed observations of regions of 
particular interest, data transmission to ground via X-band (4.8 Gb/orbit, 
Alcantara and Malindi ground stations), and prompt transmission of GRB / transient 
triggers.

\end{abstract}


\bodymatter

\section{Scientific Objectives}

The proposed GAME mission has two main scientific objectives:

a) measuring the photon spectrum and timing of the prompt emission of gamma ray 
bursts (GRBs) over a broad energy band, from 1 keV to 20 MeV, combined with arcmin 
location accuracy;

b) monitoring the X-ray sky in the 1 to 200 keV band with a few arcmin source 
location accuracy and a few mCrab daily sensitivity over a large field of view (FOV).

\subsection{Gamma-ray bursts}

Discovered in the late 1960s by military satellites and revealed to the scientific 
community in 1973, GRBs are one of the most intriguing ``mysteries" for modern 
science \cite{Meszaros06,Gehrels09,Zhang14}. Indeed, despite 
the fact that % bob edit
they are very 
bright (fluences up to more than 10$^{-4}$ erg cm$^{-2}$ 
released in a few tens of s) and very 
frequent (about 0.8/day as measured by low Earth orbit satellites) phenomena, their origin and 
the physics at the basis of their complex emission remained mostly obscure for more 
than 20 years. 
This remains true even today, %And even today, 
despite the huge observational efforts of the 
past %last 
decades, which have provided, among other 
things, 
1) a good characterization of the bursts 
temporal and spectral properties, 
2) the accurate localization and discovery of 
their multi-wavelength afterglow emission by BeppoSAX, 
3) the determination of 
their cosmological distance scale and the evidence of a connection with peculiar 
type Ib/c SNe, our understanding of the GRB phenomenon is still affected by several 
relevant open issues, both from the observational and theoretical points of view. 
Many of these 
questions 
can be addressed only by broad band sensitive measurements of the 
prompt emission, from several MeVs down to $\sim$1 keV. In particular, the extension of 
the energy band down to soft X-rays is of key importance for testing prompt 
emission models, for studying the properties of the circum-burst environment from 
the detection of transient absorption features and/or variable NH, for the 
investigation of X-ray Flashes (XRFs, which may constitute the bulk of the true GRB 
population), 
and 
for increasing the detection rate of high redshift GRBs (of paramount 
importance for the study of the early Universe). It is also fundamental, for the 
advancement of GRB science, to have a devoted mission in the $>$2020 time-frame 
in order to continue the alert, presently provided by the Swift satellite, of space 
and ground multi-wavelength telescopes.

GAME will give the answer to still 
other %several 
open questions of fundamental 
importance for the physics of the GRB phenomenon and for exploiting GRBs as 
cosmological rulers, among which 
are:

\begin{itemize}

\item to detect expected transient X-ray absorption column and absorption features 
for tens of medium/bright GRBs per year. These 
measurements are of paramount importance for the understanding of the properties of 
the Circum-Burst Matter (CBM) and hence the nature of GRB progenitors (a still 
fundamental open issue in the field). In addition, the detection of transient 
features\cite{Amati00} can allow the determination of the GRB redshift to be compared, 
when it is 
the case, with that determined from the optical/NIR lines.

\item to perform unbiased measurements of time resolved spectra within single GRBs 
down to about 1 keV. This is crucial for testing models\cite{Ghirlanda07,Frontera13}
of GRB prompt emission 
(still to be settled despite the considerable amount of observations).

\item to provide a substantial increase with respect to the past and current 
missions) in the detection rate of X-Ray Flashes (XRF), a sub-class of soft / 
ultra-soft events which could constitute the bulk of the GRB population and could 
be the missing link between high luminosity hard bursts and the low 
luminosity/relatively soft GRBs with associated SN events\cite{Amati04,Sakamoto05}.

\item to significantly increase the GRB detection up to very high redshift (z $>$ 8), 
which is of fundamental importance for the study of evolutionary effects, the 
tracing of the star formation rate, ISM evolution, and possibly unveiling 
population III stars\cite{Tanvir09}. 

\item to perform an accurate determination of spectral peak 
energy, which is a fundamental quantity for the test and study of spectrum-energy 
correlations and the possible use of GRBs as cosmological 
probes\cite{Ghirlanda04,Amati08,Amati13}.

\item to provide fast (within 1 min) and accurate (within 1--2 arcmin) location of 
the detected GRBs to allow their prompt multi-wavelength follow-up with ground and 
space telescopes, thus leading to the identification of the optical counterparts 
and/or host galaxies and to the estimate of the redshift, a fundamental measurement 
for the scientific goals listed above.

\end{itemize}

We remark that GRB science is of interest to several fields of modern astrophysics 
and cosmology, such as: physics of matter under extreme conditions, core-collapse 
and black-hole formation in massive stars, peculiar SNe, star formation rate 
evolution up to the early Universe, first generation of stars, measurement of the 
geometry and expansion rate of the Universe. These topics fit very well the ESA Cosmic 
Vision 2015--2025 plan, in particular, themes 3.3 (``Matter under extreme 
conditions"), 4.3 (``The evolving violent Universe") and 4.1 (``The early Universe") 
and of the recommendations of the ASTRONET Science Vision and Road-map (in 
particular, Theme 2: ``Do we understand the extremes of the Universe?")

\subsection{X-ray all-sky monitoring}

Besides GRB science, an instrumentation with a large FOV (order of a few sr), an 
energy band extending down to 1 keV, and a source location capability in the arcmin 
range, can also be devoted to X-ray all-sky monitoring observations. As stressed 
also in the ASTRONET Science Vision and Road map, the wide-field X-ray monitoring is 
a crucial task for X-ray Astronomy, due to the large variability of almost all 
classes of sources. However, the most sensitive observatories have in general a 
narrow FOV ($\sim$1$^\circ$ or less) and are designed to perform studies of individual 
sources. Instead the perspectives of wide field monitoring are not satisfactory: 
RXTE/ASM completed its mission operations and ISS/MAXI will likely operate 
for the next 2--4 years. In addition to the sky survey, we plan to also perform 
follow-up broadband (1--200 keV) observations of transient sources in outbursts 
previously discovered with the scanning mode. Also, deep observations of persistent 
but variable sources are foreseen for studying their time behavior and their 
spectral evolution. We also expect to trigger many TOO multi-wavelength 
observations (from radio to optical to X-rays). Most of the sources that will be 
detected and studied through X-ray all-sky monitoring provide unique insights on 
the properties of neutron stars and black-holes and the physics of matter accretion 
onto these objects. Thus, this science, in addition to the ASTRONET 
recommendations, fits very well the Cosmic Vision 2015--2025 plan themes 
``Matter under extreme conditions" (3.3) and ``The evolving violent Universe" (4.3).

Concerning the all-sky monitoring, the scientific objectives of GAME include:

\begin{itemize}

\item detection and localization within a few arcmin of Soft Gamma-ray Repeaters 
(SGR), X-ray bursts (XRB) and many other classes of galactic X-Ray Transients 
(XRT), like, e.g., Galactic low and high mass X-ray binaries in outburst, 
cataclysmic variables, accreting ms pulsars, etc., for spectral and timing studies;

\item to trigger follow-up observations by ground and space observatories, a 
fundamental service for the world-wide community;

\item to perform an all-sky survey in the hard X-ray band up to 200 keV, 
contemporary to that of eROSITA at lower energies.

\end{itemize}

The above science goals address fundamental questions of the ESA Cosmic Vision 
2015--2025, like the following: what are the fundamental laws of the Universe? What 
is the physics of matter under extreme conditions? How did the Universe originate and 
what is made of? GRBs not only allow us to investigate the physics of the most 
energetic phenomena, but are expected to be become, with the mission we are 
proposing, a new well established probe of cosmology theories.

\section{Scientific Requirements}

The scientific goals we have discussed in the previous section define the 
instrumentation and mission profile requirements.

\subsection{Payload requirements}

The scientific requirements for the instrumentation can be summarized as follows.

\begin{itemize}

\item Energy pass-band: for GRBs, from $\sim$1 keV up to $\sim$20 MeV. 
The lower threshold 
is fundamental for the 
study of transient X-ray absorption features and the substantial increase in the 
detection and study of XRFs and high-z GRBs. The broad band is of key importance 
for the identification estimate of fundamental spectral parameters.
For all sky X-ray monitoring, from $\sim$1 keV up to $\sim$200 keV. The extension to the 
hard X-rays with good efficiency optimizes the spectral sensitivity to the hardest 
X-ray transients and SGRs.

\item Energy resolution: 300 eV for photon energies $<$10 keV. It is required for 
the study of the 
expected absorption features from GRBs. An energy resolution $<$15\% for photon 
energies $>$50 keV is enough for accurate measurement of the peak energy of the  
$\nu$F$\nu$ spectrum.

\item Source location accuracy: a few arcmin. 
This % It 
is required to trigger and allow 
follow-up observations of the detected transients by other telescopes, but it is 
also essential for fulfilling the GRB scientific objectives, the all-sky monitor 
functionality and the X-ray all-sky survey science objectives.

\item Field of View (FOV): about 4 sr (Partially Coded Field of View, PCFOV) 
combined with a sensitivity of $\sim$500 mCrab (in 1s integration time) in the 3-50 
keV energy band. 
This %It 
is required in order to allow detection and localization of 
$\sim$150 GRBs and XRFs per year (including $\sim$2--4 events at z $>$ 6), to perform a 
sensitive time resolved spectroscopy of $\sim$2/3 of them and to use the instrument 
as an all-sky monitor for Galactic transients (SGRs and other XRTs).

\item Average effective area: $\sim$1000 cm$^2$ up to 200 keV, $\sim$500 cm$^2$ 
up to $\sim$20 MeV within a FOV of 2 sr. 
This %It 
is required to allow sensitive broad 
band spectroscopy of GRBs (for GRB physics and cosmology studies) and to increase 
the overall trigger efficiency, especially for short (and spectrally hard) GRBs.

\item On board data handling electronics. 
This %it 
is required to identify GRB trigger, 
discrimination of false triggers and fast source position reconstruction in order 
to allow prompt alert distribution and follow-up observations.

\end{itemize}

\subsection{Mission profile requirements}

The scientific objectives of GAME require a low and stable background level, thus a 
flight in a nearly equatorial low Earth orbit ($\sim$600 km) with a desirable 
inclination lower than 5$^\circ$. This also allows 
the reduction of %to reduce the
radiation damage to the 
detectors and ASICs. Also, %for  
the achievement of the scientific objectives and %to 
providing prompt triggers %, it is also required to have 
requires
the capability of promptly 
transmitting to the ground alerts with coordinates and basic information of at least 
the detected GRBs. Compatibly with 
solar %Sun 
and thus thermal constraints, the 
orientation of the spacecraft is required to be continuously drifting to cover in 
each orbit a large fraction of the sky. For particularly interesting sky fields 
pointed observations will be also required.

\begin{figure}[t!]
\psfig{file=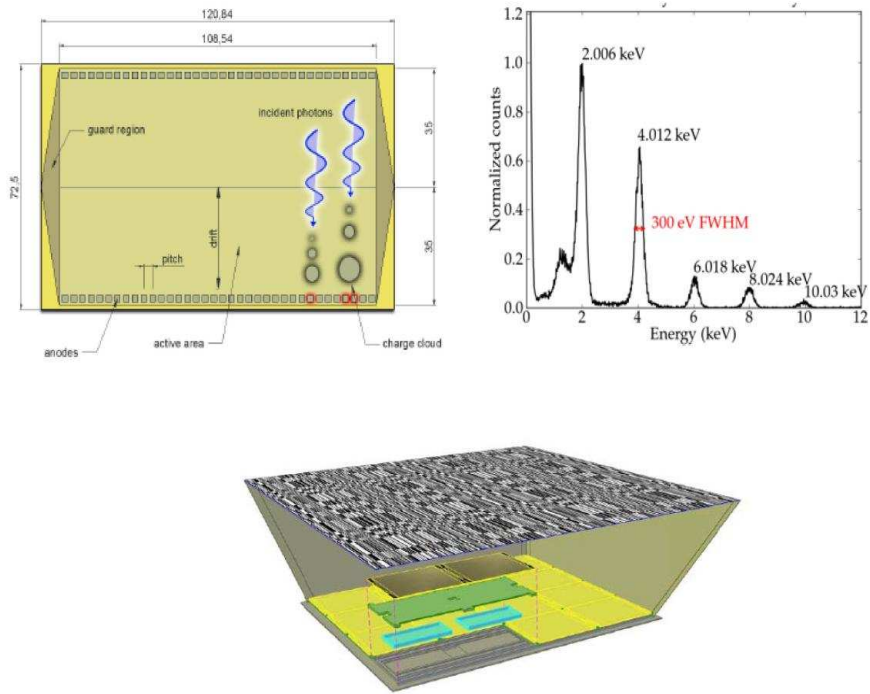,width=12cm}
\vspace{-5cm}
\caption{X-Ray Monitor (XRM): sketch of the detector's working principle (top left),
spectroscopic performance (top right) and sketch of a single camera assembly.}
\label{fig1}
\end{figure}

\section{The Proposed GAME Payload}

The approach is to optimize instrumentation for both GRB prompt emission study and 
the all-sky X-ray monitoring. For the GRB study, we propose a broad band (1--20000 
keV) GRB monitor with an imaging capability at low energies, while for the ASM we 
propose an imaging capability in the entire band (1--50 keV) in which we want to 
work, with an extension up to 200 keV in a narrower FOV. The instrumentation 
proposed consists of an X-Ray Monitor, XRM (Fig.~1), made of a set of 
9 pairs of Si based 
independent imaging cameras (1--50 keV). Each pair of cameras consists of a wide 
field Silicon Drift detectors (initially developed for the ALICE experiment for the 
CERN/LHC accelerator and later adapted for space astronomy, e.g., for 
the
LOFT 
mission\cite{Feroci12}), surmounted by orthogonally located 1D 
coded masks, to get, for each pair, 
a bi-dimensional (2D) imaging. Instead the extension of the monitoring at hard 
X-ray energies is obtained with a CZT position sensitive detector surmounted by a 2D 
coded mask working in the 10--200 keV energy range\cite{Braga06} 
(Hard X-ray Imager, HXI, Fig.~2). 
The high energy portion of 
the GRB spectrum (20--20000 keV) is measured with 2 modules of 4 NaI(Tl)/CsI(Na) 
scintillator units in phoswich configuration\cite{Frontera97} 
(Soft Gamma-ray Spectrometer, SGS; Fig.~3). 
A possible overall payload configuration is illustrated in Fig.~4.

\begin{figure}[t!]
\vspace{-7cm}
\psfig{file=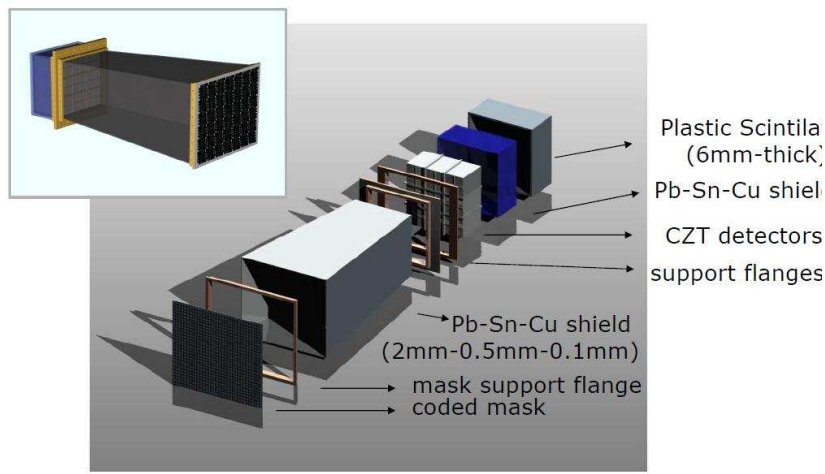,width=11cm}
\caption{Sketch of the Hard X-ray Imager (HXI).}
\label{fig2}
\end{figure}

\begin{figure}[t!]
\vspace{-5cm}
\psfig{file=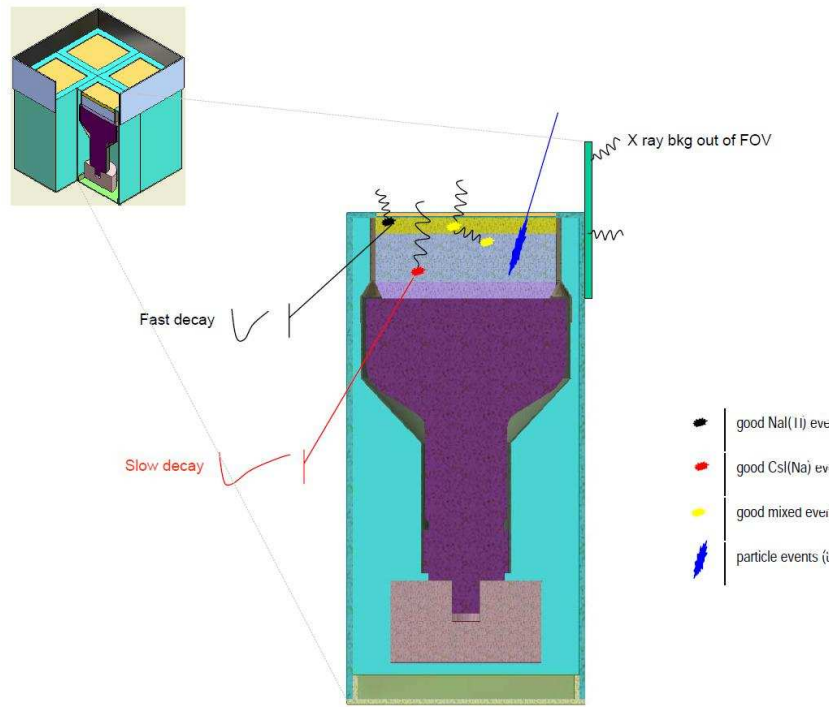,width=11cm}
\caption{Soft Gamma-ray Spectrometer (SGS): sketch of a single phoswich unit
assembly and working principle and, in top left position, sketch of 4 units
assembly.}
\label{fig3}
\end{figure}

\subsection{Instrument conceptual design and key characteristics}

XRM is a set of Si-detectors and coded-mask X-ray cameras, 3 pairs for each sky 
direction. In each camera the detection plane is made of 2x2 Silicon Drift Detector 
(SDD) tiles. In a tile (active area $\sim$7x7 cm$^{-2}$, 450$\mu$ thickness) a cloud of 
electrons generated by interaction of the X-ray photon with Si is drifted toward 
the read out anodes driven by a constant electric field\cite{Campana11}. The 
Si tile is electrically divided in two halves with 2 series of about 350 anodes 
(pitch $\sim$200$\mu$) at the two edges and the highest voltage along the symmetry axis. 
The requirement of fine pitch, small parasitic capacitance and low power 
consumption require a high density read-out system based on ASIC (Application 
Specific Integrated Circuit) that includes several independent, low noise complete 
spectroscopic chain made of charge preamplifier, pulse shaper and amplifier, 
discriminator, etc. The detection plane of each camera is surmounted by 1D coded 
mask placed at 20 cm distance. 

The SGS is composed by 8 scintillator units, each one being an independent detector 
in which the active part is made by 1 thick cm thick NaI(Tl) (top) and 4 cm CsI(Na) 
(bottom) scintillators both viewed by a single Photomultiplier Tube (PMT) (phoswich 
configuration). The useful cross section is 14x14 cm$^{-2}$. The PMT signal is analyzed 
in both shape and amplitude in order to measure the energy losses separately in 
NaI(Tl) and CsI(Na). In the low energy band (20--200 keV) of SGS operation, the 
CsI(Na) will have the role of actively shielding and thus rejecting the background 
(BKG) coming from the bottom (e.g., terrestrial albedo), while in the 100-20000 
keV, the CsI(Na) will act as main detectors. In this energy band also the losses in 
both crystal materials will be analyzed. Each set of 4 phoswich units will be 
passively collimated to further reduce the diffuse background. Each set of 4 
phoswich units will be offset by 10$^\circ$ with respect to the HXI axis to cover most of 
the FOV of the XRM. A gain control system based on the SAX/PDS design will be 
implemented. A ground calibration campaign will be performed for each instrument 
before flight, using the already available X-ray facilities in the GAME consortium 
(e.g., the LARIX facility at the University of Ferrara, already used for the 
INTEGRAL/JEM-X ground calibration). 
The main scientific characteristics of the GAME instruments, together with a 
summary of the required resources, are reported in Table 1.

\begin{table}[t!]
\tbl{Main characteristics and resources of the GAME instruments.}
{\psfig{file=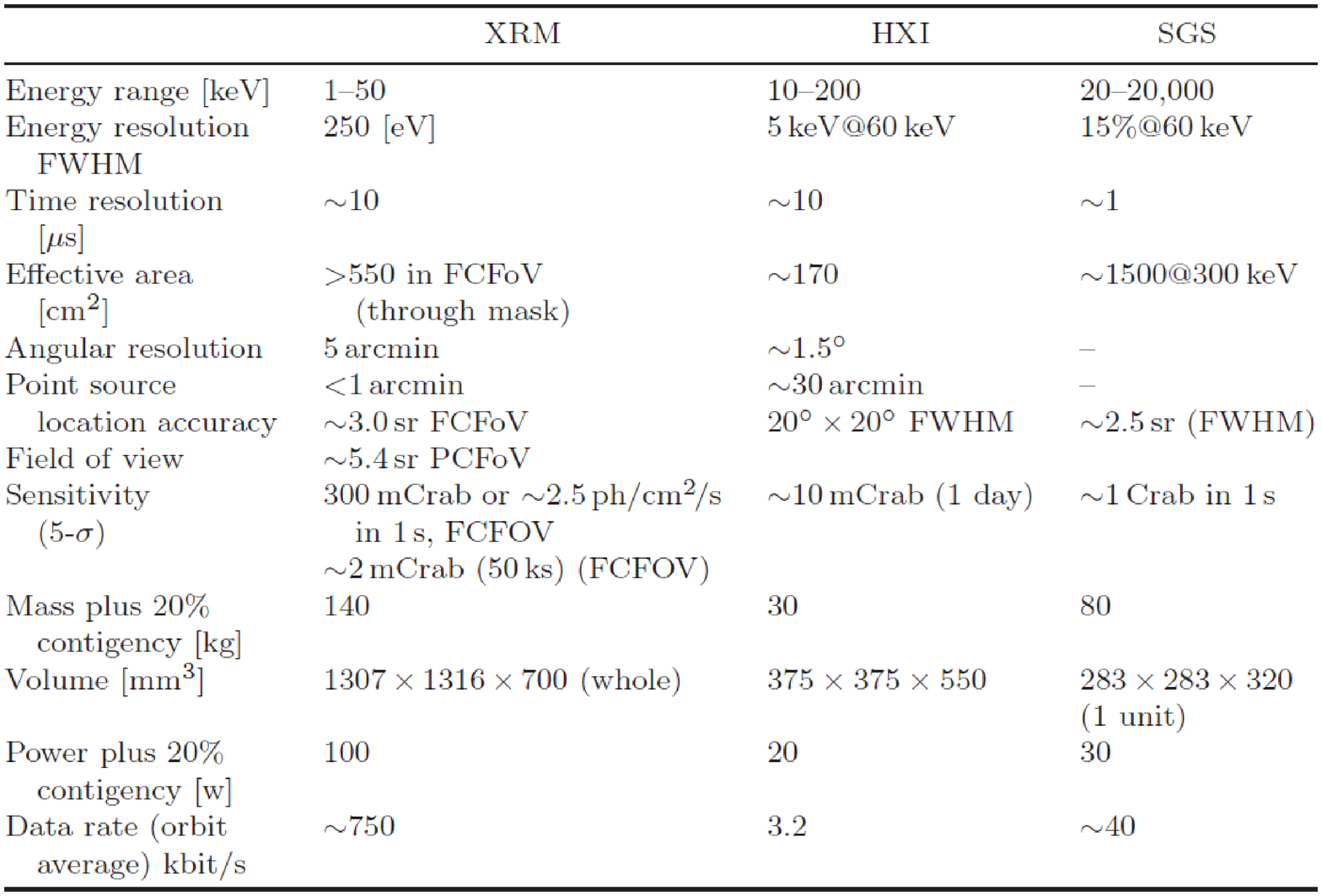,width=14cm}}
\label{tab1}
\end{table}

\begin{figure}[t!]
\vspace{-13cm}
\psfig{file=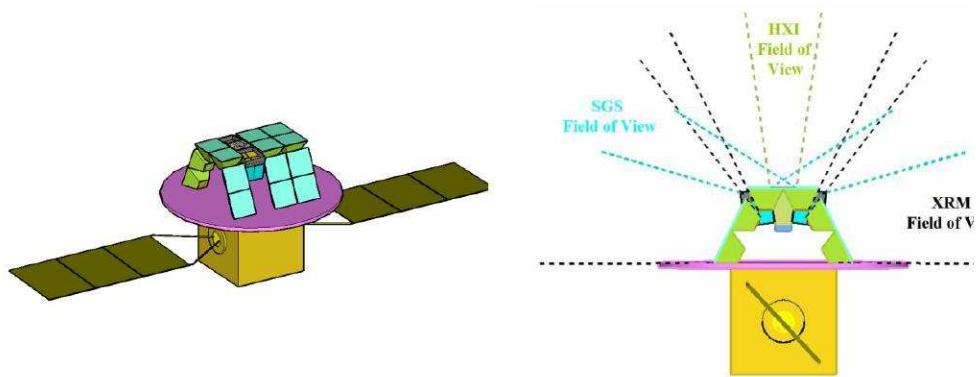,width=13cm}
\caption{An hypothesis of allocation of XRM, HXI and SGS. The dimensions of 
the bus are $\sim$1 m$^3$.}
\label{fig4}
\end{figure}

\begin{table}[t]
\tbl{Main characteristics of the whole satellite.}
{
\psfig{file=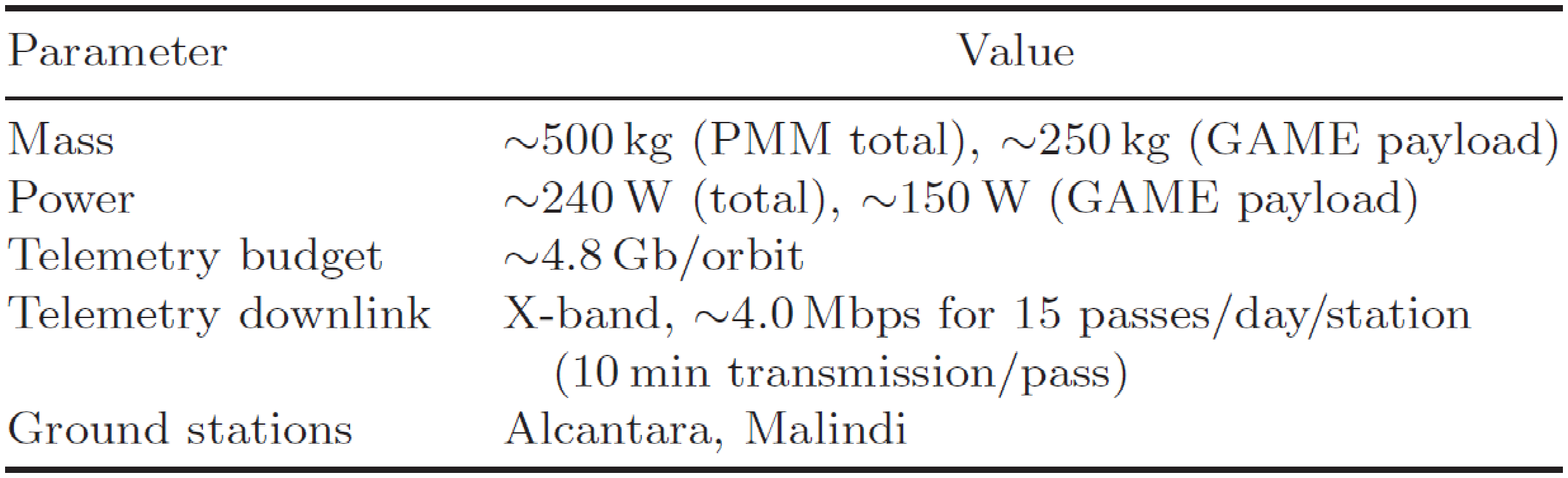,width=13cm}}
\label{tab2}
\end{table}

\begin{figure}[t!]
\vspace{-8cm}
\centerline{\psfig{file=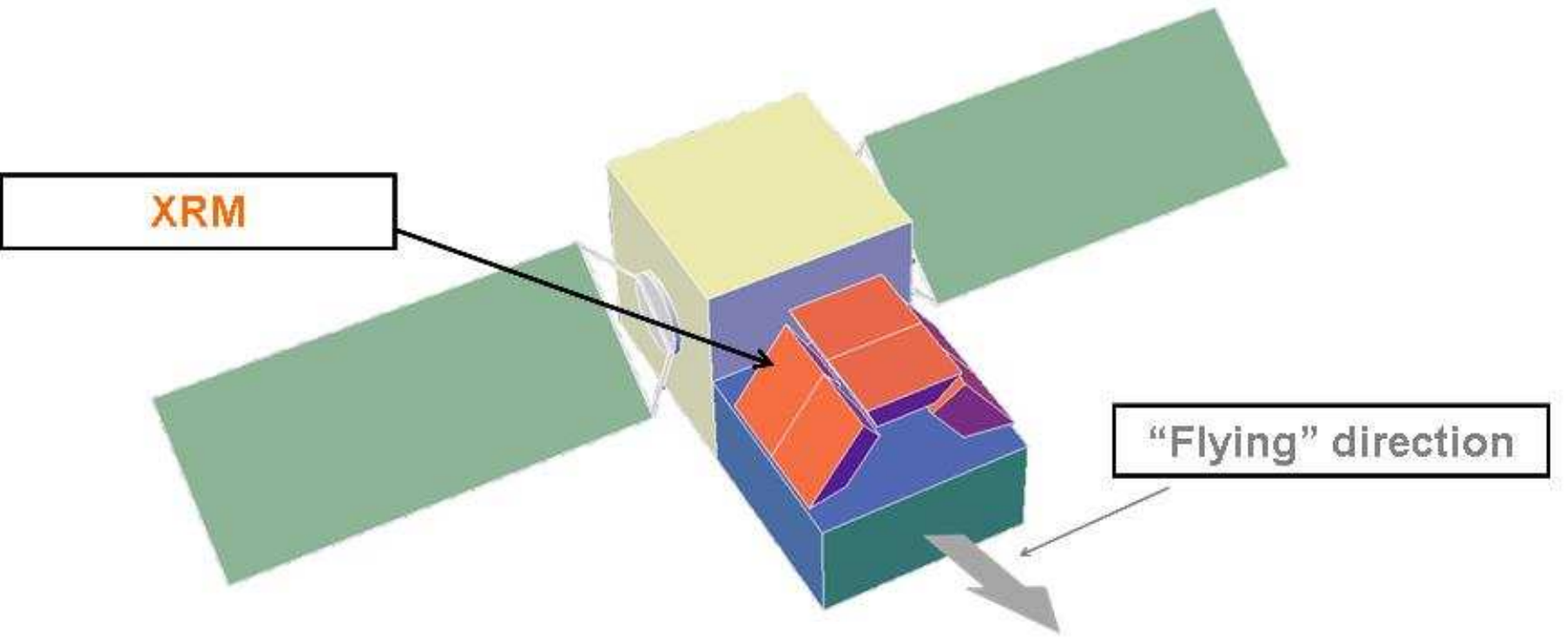,width=8cm}\psfig{file=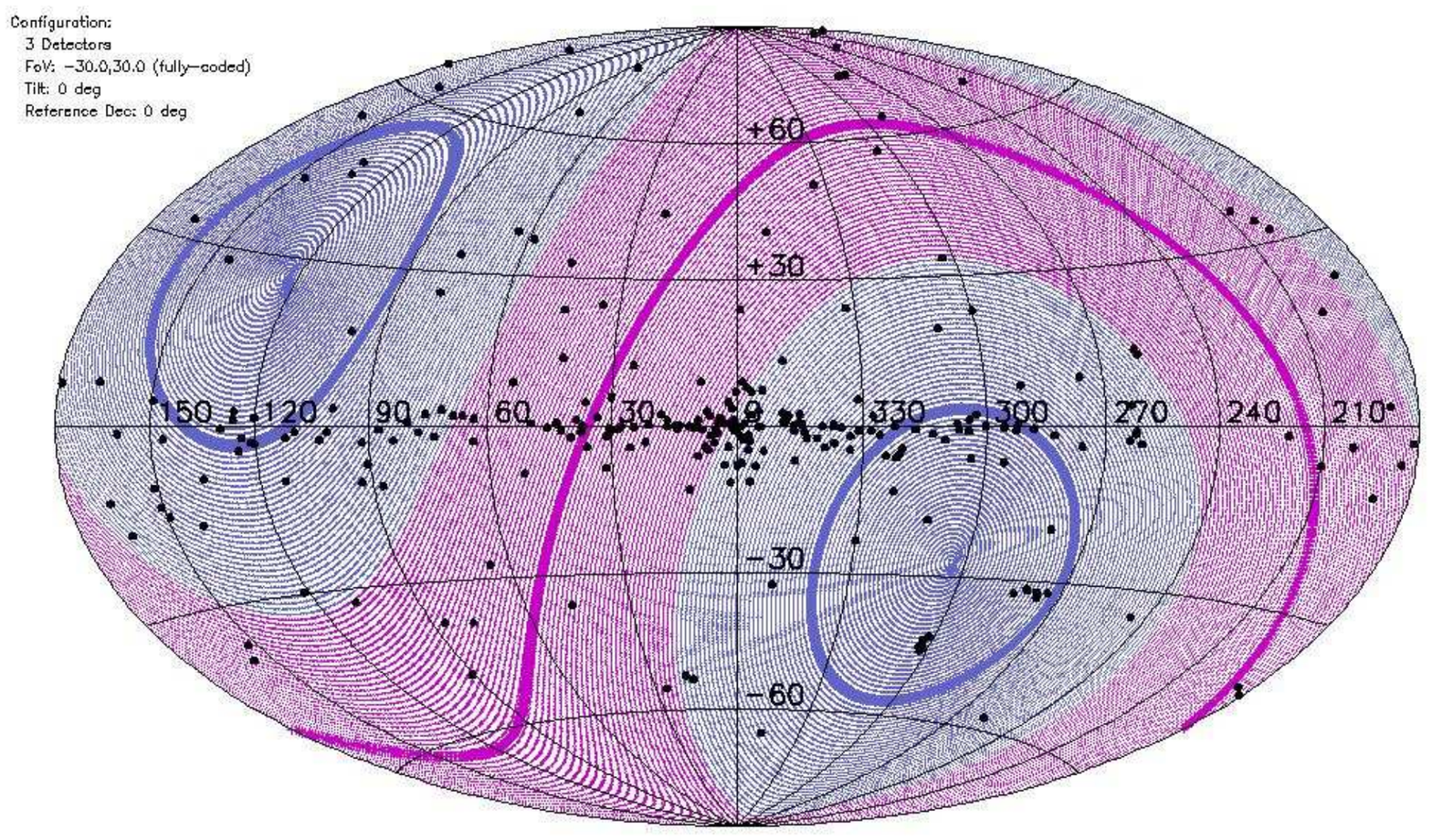,width=8cm}}
\caption{Left: flying configuration and pointing strategy of the satellite (for clarity, only 6 units
of the XRM are shown). Right: XRM sky coverage (every orbit).}
\label{fig5}
\end{figure}

\section{Mission Profile}

Both the orbit and the payload volume and mass budget are compatible with the Vega 
launcher. Although the satellite is not compatible with the Vega piggy-back 
approach, its compactness and reduced size and 
weight, may make possible a dual launch approach if also the second satellite 
manifest interest as Vega piggy-back.
The scientific objectives of GAME require a low and stable background level, thus a 
flight in a nearly equatorial low Earth ($\sim$600 km, $<$ 5$^\circ$) orbit.
The observation strategy has been defined combining: 1) scientific goals; 2) 
operational constraints imposed by the payload (e.g., Sun illumination of the 
instrument FOV is not allowed); 3) simplest mission management; 4) efficient use of 
platform capability.
Consequently two operative modes have been envisaged:
scanning mode
and pointing mode. % missing period
The scanning mode is the default observational mode. It consists 
of %in 
continuously 
orienting the spacecraft, in such a way that the negative direction of the platform 
axis on which the instruments are located (Fig.~5), compatibly with the Sun 
constraints, will continuously point at the Earth center. In this way a large part 
of the sky will be observed in one orbit with XRM and SGS. When the Sun prevents us 
from performing %to perform 
the scanning strategy, the orientation of the satellite will drift to 
directions that prevent a Sun illumination of the instrument FOV keeping constantly 
a null Sun aspect angle. During these time periods, in addition to the sky 
observations, the Earth will be observed with the possibility of detecting and 
localizing Terrestrial Gamma-ray Flashes (TGFs). These are very spiky events whose 
emission physics is still 
the
subject of discussion and are possible probes of 
terrestrial thunderstorms\cite{Marisaldi10}.
%Pointing mode.\\ 
Pointed observations of particularly interesting regions previously 
scanned or sky regions requested for TOO observations will be also performed. In 
this case the sky accessible field is reduced during orbit daylight period by 
zero %null 
Sun aspect angle.
The expected mission lifetime is
4 years, extendible to further 3 years.
The foreseen telemetry budget with no data compression is compatible with the use 
of an X-band transmitter.
For the prompt transmission to ground of coordinates and basic 
information of detected GRBs or X-ray transients, a real time link with Earth is 
foreseen. A possibility we are evaluating is an innovative telemetry system, 
STRADIUM, realized by an Italian team in collaboration with the Italian Space 
Agency. It provides a near real time, bi-directional, continuous 
telemetry/telecommands link. It exploits the IRIDIUM network\cite{Ronchi08}.

The main parameters of the entire satellite are reported in Table 2. 
The Brazilian PMM 
platform allows an absolute pointing accuracy of 3 arcmin 
(3$\sigma$) and an attitude determination of 0.3 arcmin (3$\sigma$). 
The XRM and the HXI require their 
pointing direction to be known better than 1 arcmin in order to correctly 
reconstruct their images. A pre-launch optical alignment with the star trackers 
will ensure this. The baseline is a Vega launch devoted to GAME. An alternative 
launch option may include the use of Vega in full piggy-back configuration, which 
is compatible with the mass and volumes of the proposed bus and payload.

\begin{figure}[t!]
\vspace{-8cm}
\centerline{
\hspace{4cm}
\psfig{file=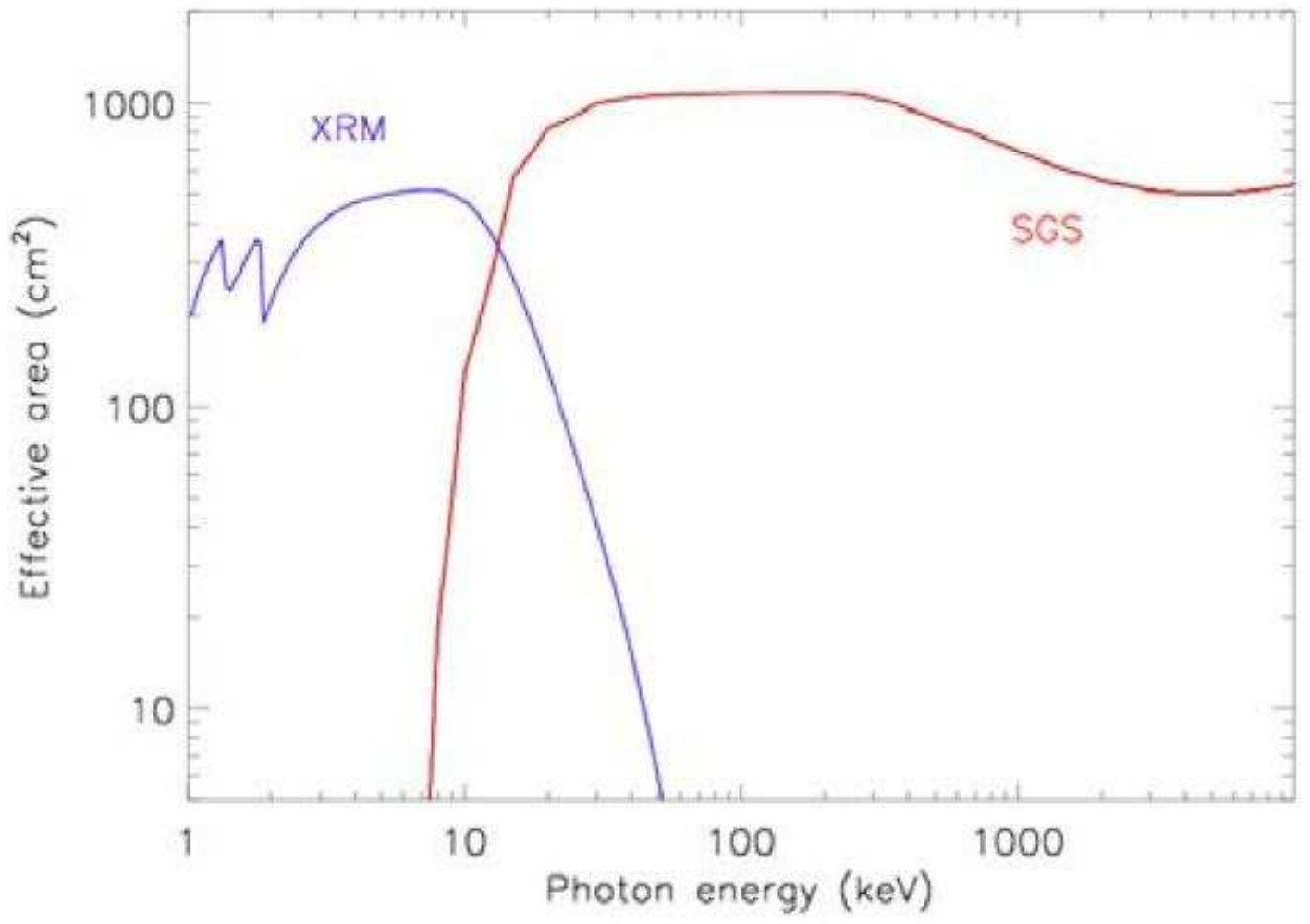,width=9.5cm}
\hspace{-2.7cm}
\psfig{file=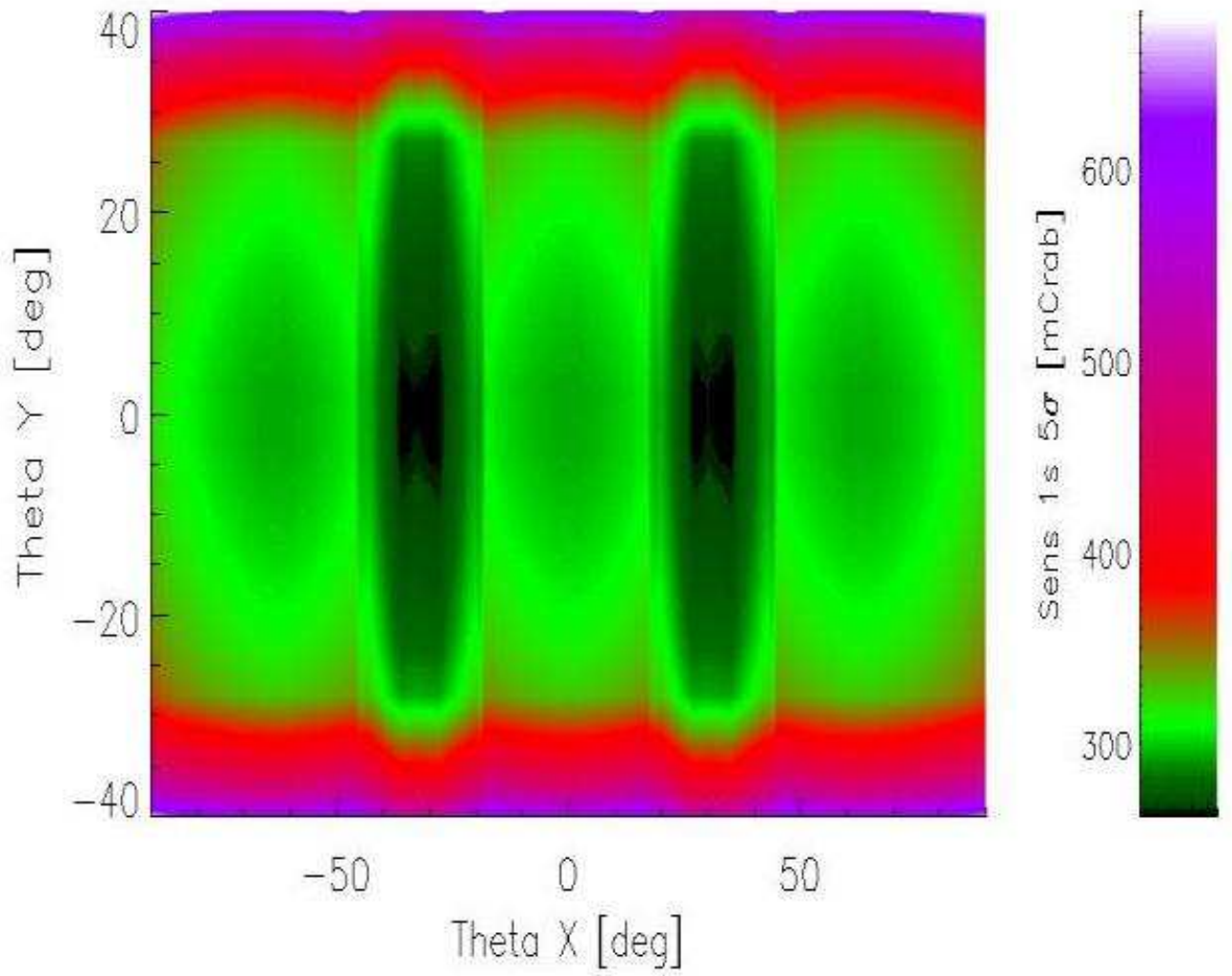,width=10cm}}
\caption{Effective area (left) and sensitivity (5$\sigma$, 1s; right) within the 
FOV of the XRM.}
\label{aba:fig6}
\end{figure}

\begin{figure}[t!]
\centerline{\psfig{file=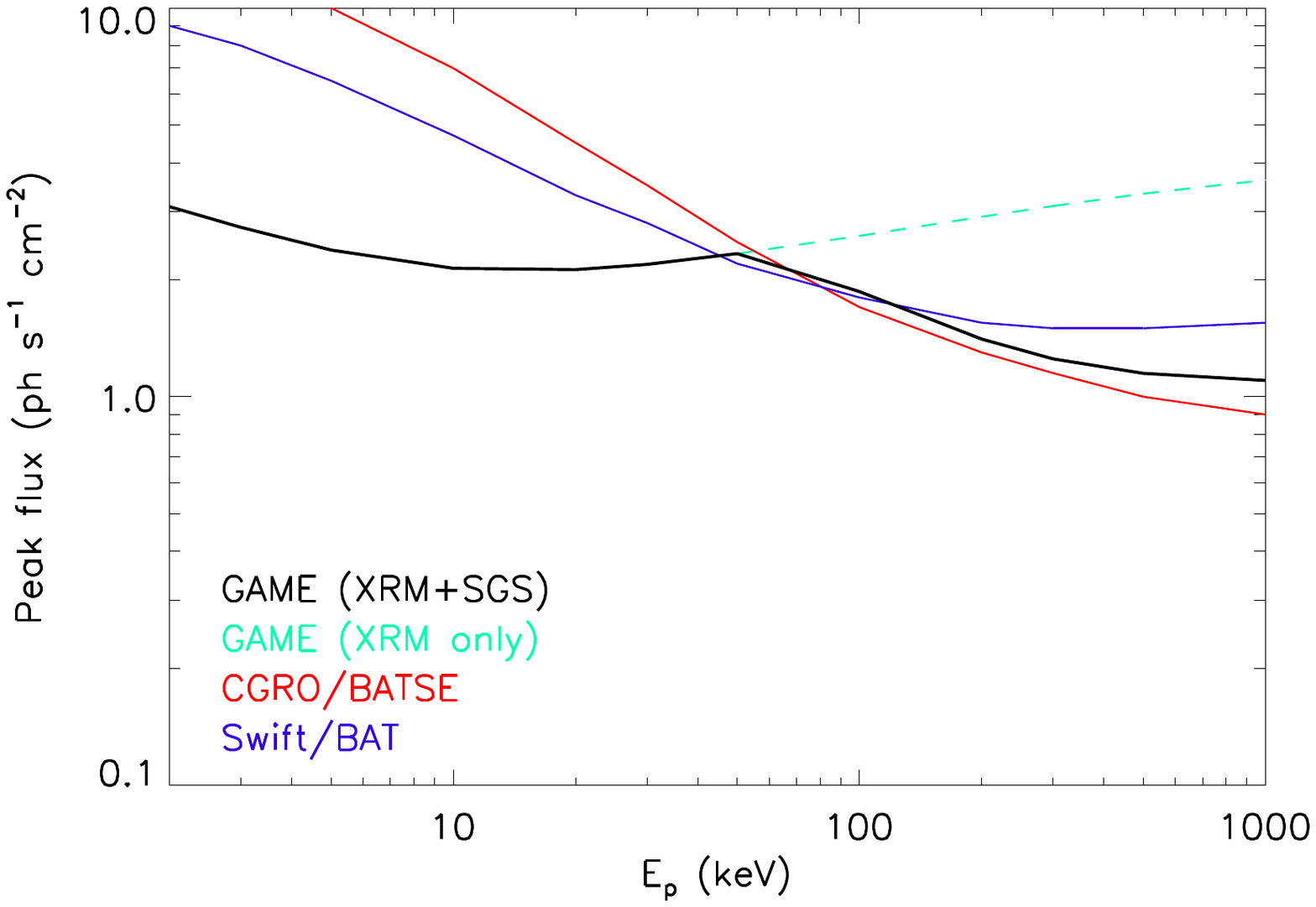,width=8cm}\hspace{-1cm}\psfig{file=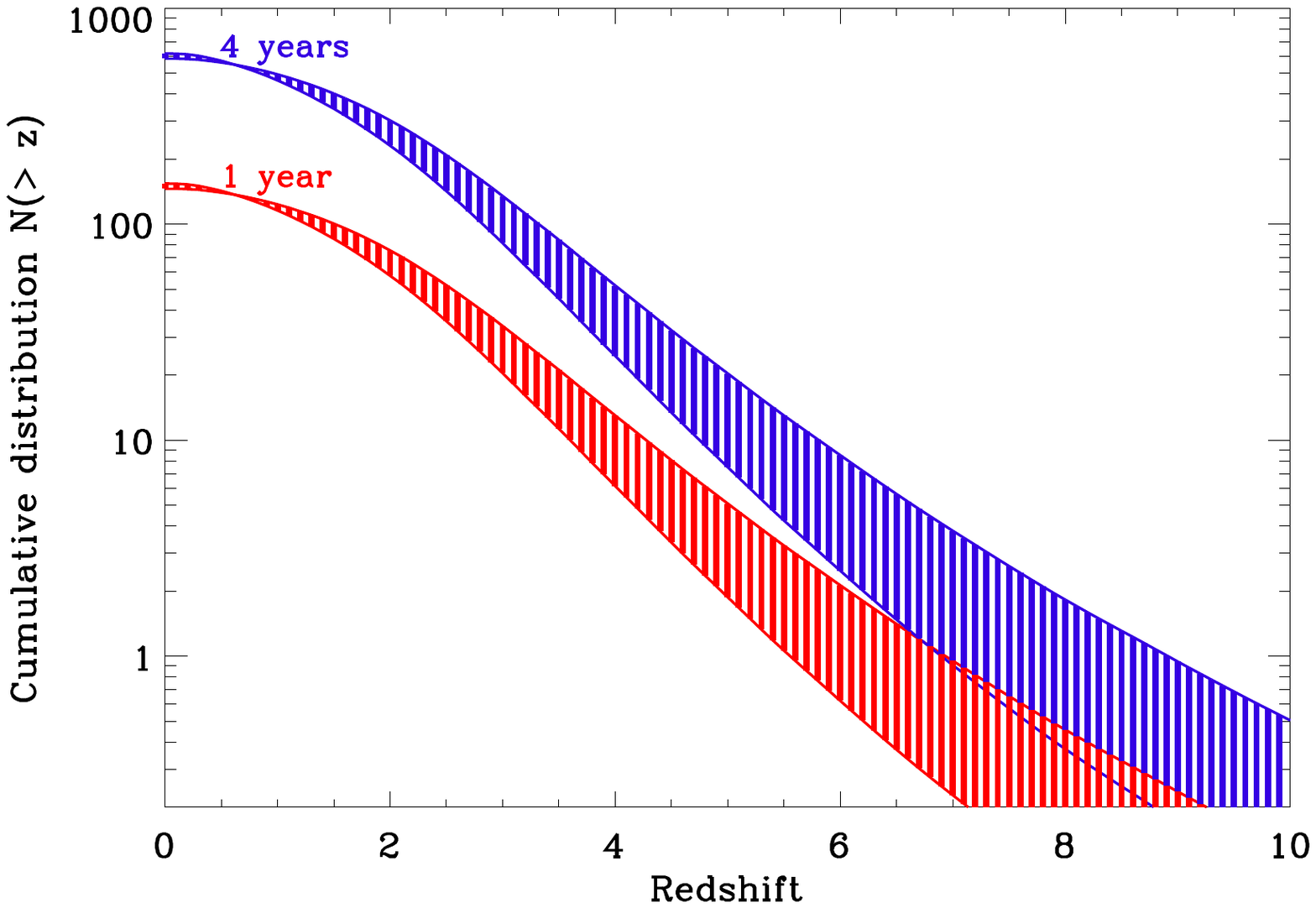,width=8cm}}
\vspace{-4cm}
\caption{Left panel (adapted from a figure by Ref.~\refcite{Band03}): expected 
GAME flux sensitivity as a function of GRB spectral peak energy (Ep) compared with 
that of BATSE and Swift/BAT. Right panel: cumulative redshift distribution of 
long GRBs predicted for GAME. Shaded regions take into account the error on the evolution parameters: the red (blue) shaded region refers to 1yr (4yr) of mission.}
\label{aba:fig7}
\end{figure}

\begin{figure}
\vspace{-11cm}
\hspace{2cm}
\centerline{
\hspace{5cm}
\psfig{file=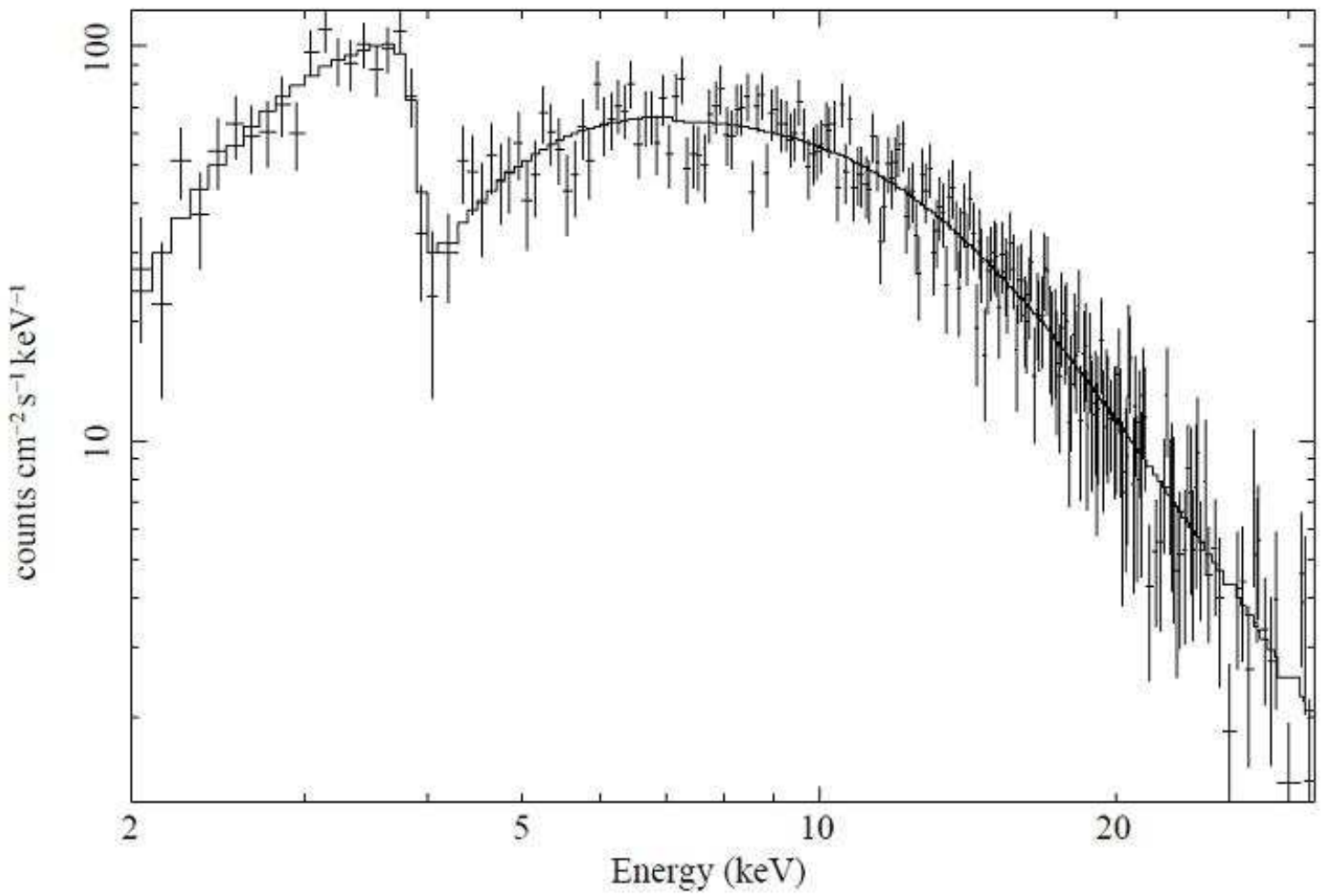,width=9.5cm}
\hspace{-2.5cm}
\psfig{file=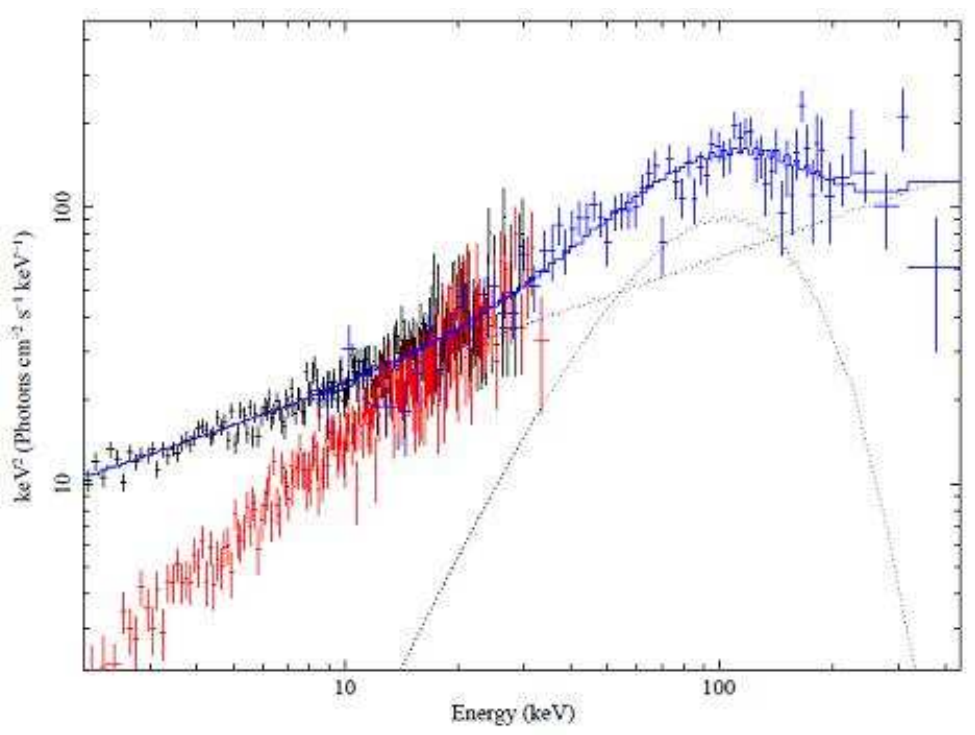,width=12cm}}
\caption{Left: the X-ray transient absorption feature observed with BeppoSAX/WFC in the 
first 13 s of GRB 990705\cite{Amati00} would be detected by GAME with a significance 
larger than 10$\sigma$. Right: simulated XRM spectra of the first $\sim$50s of 
GRB\,090618
obtained by
assuming either the Band function (black) or the power-law
plus black-body model (red) which equally fit\cite{Izzo12} the Fermi/GBM 
measured spectrum,
which is also shown (blue).
The black-body plus power-law model components best-fitting the
Fermi/GBM spectrum are also shown (black dashed lines).} 
\label{fig8}
\end{figure}

\section{Expected Performances}

Figures 6 to 8 show examples of the expected scientific performances of 
GAME. In Fig. 6 we show the effective area as a function of energy and the 
5$\sigma$ sensitivity in 1s observation 
time of the XRM in its FOV. As can be seen, we have at least 500 cm$^{-2}$ 
in most of its 
FOV. In Fig.~7 left, we show the expected GAME flux sensitivity as a function of 
GRB spectral peak energy (Ep) compared with that of BATSE and Swift/BAT, while in 
Fig.~7 right we show the cumulative redshift distribution of long GRBs predicted for 
GAME. This distribution has been obtained assuming a broken power-law GRB 
luminosity function and a pure density evolution (see Ref.~\refcite{Salvaterra12}
for 
the details of the model). The model's free parameters has been 
determined by %obtained 
fitting the 
observed differential number counts and the observed redshift distribution of a 
redshift complete sample of bright Swift GRBs\cite{Salvaterra12}. Without any 
change in 
the model's free parameters, 
the same model is able to reproduce the redshift 
distribution of GRBs at the Swift sensitivity as collected by GROND. Shaded regions 
take into account the error on the evolution parameters: the red (blue) shaded 
region refers to 1yr (4yr) of mission. At the flux limit of GAME we predict up to 
16 GRBs to lie at z$>$6 in the full 4 yr mission life. For a comparison, in 7 
years of operations, only 4 GRBs detected by Swift have a measured redshift higher 
than 6.
In Fig.~8 (left) we compare the X-ray transient absorption feature at 3.8 keV observed 
with BeppoSAX/WFC in the first 13s of GRB\,990705\cite{Amati00}
and that expected if that feature would be observed with 
GAME. The 3.8 keV edge would be detected with a significance of more than 12$\sigma$ 
against the $\sim$4$\sigma$ significance obtained with SAX/WFC.
In Fig.~8 (right) we show the
simulated WFM spectra of the first $\sim$50s of GRB\,090618
obtained by
assuming either the Band function or the power-law
plus black-body model (red) which equally fit the Fermi/GBM measured 
spectrum\cite{Izzo12},
which is also shown.
As can be seen, only thanks to the
extension of the spectral measurement below 10 keV allowed by the XRM, it is
possible to discriminate between the Band and black-body plus power-law models.

\end{document}